\pdfoutput=1 
\RequirePackage{build-config}
\newcommand{\formattype}{\formattypeIEEE}
\newcommand{\whichExperiment}{apr-10-more-combos}
\newcommand{\whichVulnExperiment}{federico-vuln-exp}

\RequirePackage{ifthen}

\newcommand{\formattypeIEEE}{formattypeIEEE}

\ifisdraft
  \ifthenelse{\equal{\formattype}{\formattypeIEEE}}{
    \documentclass[10pt,conference]{IEEEtran}
    \IEEEoverridecommandlockouts
  } {
    \ifisarxiv
      \ifisanonymous
        \documentclass[acmsmall,screen,nonacm,review,anonymous]{acmart/acmart}
        \settopmatter{printfolios=true,printccs=false,printacmref=false}
      \else
        \documentclass[acmsmall,screen,nonacm,review]{acmart/acmart}
        \settopmatter{printfolios=true,printccs=false,printacmref=false}
      \fi
    \else
      \ifisanonymous
        \documentclass[acmsmall,screen,review,anonymous]{acmart/acmart}
        \settopmatter{printfolios=true,printccs=true,printacmref=true}
      \else
        \documentclass[acmsmall,screen,review]{acmart/acmart}
        \settopmatter{printfolios=true,printccs=true,printacmref=true}
      \fi
    \fi
  }
\else
  \ifthenelse{\equal{\formattype}{\formattypeIEEE}}{
    \documentclass[10pt,conference]{IEEEtran}
    \IEEEoverridecommandlockouts
  } {
    \ifisarxiv
      \ifisanonymous
        \documentclass[acmsmall,screen,nonacm,anonymous]{acmart/acmart}
        \settopmatter{printfolios=true,printccs=false,printacmref=false}
      \else
        \documentclass[acmsmall,screen,nonacm]{acmart/acmart}
        \settopmatter{printfolios=true,printccs=false,printacmref=false}
      \fi
    \else
      \ifisanonymous
        \documentclass[acmsmall,screen,anonymous]{acmart/acmart}
        \settopmatter{printfolios=true,printccs=true,printacmref=true}
      \else
        \documentclass[acmsmall,screen]{acmart/acmart}
        \settopmatter{printfolios=true,printccs=true,printacmref=true}
      \fi
    \fi
  }
\fi

\ifthenelse{\equal{\formattype}{\formattypeIEEE}}{

} {
  \citestyle{acmauthoryear}
}

\newwrite\abstractoutput
\immediate\openout\abstractoutput=\jobname.abstract.output

\ifthenelse{\equal{\formattype}{\formattypeIEEE}}{
  \newcommand{\makeTitleAndAbstract}{
    \write\abstractoutput{\theAbstract}
    \maketitle
    \begin{abstract}
    \theAbstract
    \end{abstract}

    \begin{IEEEkeywords}
      package-management, Max-SMT, NPM, Rosette, dependency-management, JavaScript
    \end{IEEEkeywords}
  }
}{
  \newcommand{\makeTitleAndAbstract}{
    \write\abstractoutput{\theAbstract}
    \begin{abstract}
    \theAbstract
    \end{abstract}
    \maketitle
  }
}

\usepackage{paper-specific-macros}
\usepackage{package-config}

\newcommand{\dataNumNPMFailures}{47}

\newcommand{\dataMinMinNPMFailures}{26}

\newcommand{\dataMaxMinNPMFailures}{28}

\newcommand{\dataNumPIPUnsupported}{19}

\newcommand{\dataFractionPIPUnsupported}{1.9\%}

\newcommand{\dataFractionShrinking}{21\%}

\newcommand{\dataFractionNewer}{14\%}

\newcommand{\dataFractionOlder}{5\%}

\newcommand{\dataFSShrinkageQuartileFirst}{82\%}

\newcommand{\dataMeanSlowdown}{2.6s}

\newcommand{\dataMedianSlowdown}{1.6s}

\newcommand{\dataMaxSlowdown}{329s}

\newcommand{\dataMinSlowdown}{-2.3s}

\newcommand{\dataFirstQuantileSlowdown}{0.8s}

\newcommand{\dataThirdQuantileSlowdown}{2.2s}

\newcommand{\dataStdSlowdown}{13.7s}

\newcommand{\dataCveExpSuccessCount}{472}

\newcommand{\dataCveExpMaxNpmBetterCount}{235}

\newcommand{\dataCveExpMeanCveDiff}{14.75}

\newcommand{\dataCveExpMaxNpmBetterTotalPerc}{33\%}

\begin{document}

\makeTitleAndAbstract

\section{Introduction}
\label{sec:intro}

Package managers such as NPM (the de facto package manager for JavaScript) have become essential for software development.
For example, the NPM repository hosts over two million packages and serves over 43 billion downloads weekly.
The core of a package manager is its \emph{dependency solver}, and NPM's solver tries to quickly find dependencies that are recent and satisfy all version constraints.
Unfortunately, NPM uses a greedy algorithm that can duplicate dependencies and can even fail to find the most recent versions of dependencies.

Moreover, the users of NPM may have other goals that NPM does not serve.
\begin{inparaenum}
  \item Web developers care about minimizing code size to reduce page
  load times. ``Bundlers'' such as Webpack alter the packages selected by NPM to eliminate duplicates (\cref{webpack-could-be-better}).
  \item Many developers want to avoid vulnerable dependencies and several tools detect and update vulnerable dependencies, including NPM's built-in ``audit'' command~\cite{npm-audit}.
  However, the audit command is also greedy and its fixes can introduce more severe vulnerabilities (\cref{npm-audit-could-be-better}).
  \item Finally, there are semantic reasons why many packages, such as frameworks with internal state, should never have multiple
  versions installed simultaneously. However, NPM's approach to solving this,
  known as ``peer dependencies'' is brittle and causes confusion (\cref{nobody-understands-peer-dependencies}).
\end{inparaenum}

The problem with these approaches is that they are ad hoc attempts to customize and workaround NPM's solver.
Bundlers and audit tools modify solved dependencies after NPM produces its solution.
Peer dependencies  effectively disable the solver in certain cases and rely on the developer to select unsolved dependencies at the package root.
In general, NPM cannot produce a ``one-size-fits-all'' solution that satisfies the variety of goals that developers have.
Moreover, any tool that modifies the solver's solution after solving risks introducing other problems and may not compose with other tools.

Our key insight is that all these problems can be framed as instances of a more general problem:
optimal dependency solving, where the choices of optimization objectives and constraints determine which goals are prioritized.
Due to the wide-ranging goals in the NPM ecosystem, this paper argues that NPM should allow developers to \emph{customize and combine several objectives}.
For example, a developer should be able to specify policies such as ``dependencies must not have any critical vulnerabilities'', ``packages should not be duplicated'', and combine these with the basic objective of ``select the latest package versions that satisfy all constraints.''
To make this possible and evaluate its effectiveness, we present \namenpm: a complete, drop-in replacement for NPM, which empowers developers to combine multiple objectives.

The heart of \namenpm{} is a generalized model of dependency solvers that we call \namebackend{}.
\namebackend{} has a high-level DSL for specifying the syntax and semantics of package versions, version constraints, optimization objectives, and more.
Under the hood, \namebackend{} uses a solver-aided language to produce a problem for a Max-SMT solver, which ensures that its solution is optimal, unlike NPM's greedy approach.
Although we apply \namebackend{} to build a package manager for NPM, we believe that the generality of \namebackend{} will make it possible to build customizable dependency solvers for other package ecosystems as well.

We use \namenpm{} to conduct an empirical evaluation with a large dataset of widely-used packages from the NPM repository. Our evaluation shows that \namenpm{} outperforms NPM in several ways:
\begin{enumerate}
  \item chooses newer dependencies compared to using NPM for \dataFractionNewer{} of packages with at least one dependency.
  \item shrinks the footprint of \dataFractionShrinking{} of packages with at least one dependency.
  \item reduces the number or severity of security vulnerabilities in \dataCveExpMaxNpmBetterTotalPerc{} of packages with at least one dependency.
\end{enumerate}
Overall, \namenpm{} takes just \dataMeanSlowdown{} longer than NPM on average, though encounters some outliers which solve
significantly more slowly with \namenpm{}, which is reflected in the standard deviation of the slowdown (\dataStdSlowdown{}).

This paper makes the following contributions:

\begin{enumerate}

  \item \namebackend: an executable semantics for dependency solvers, which are
        the key component of package managers. \namebackend{} is parameterized along
        several key axes, which allows for customization of constraints and optimization objectives.
        

  \item \namenpm: a drop-in replacement for NPM that allows the user to
        solve dependencies with several objectives, including maximizing secure dependencies, decreasing code size, and maximizing up-to-date dependencies. \namenpm{} is implemented by instantiating \namebackend{} to use NPM's notions of versions and version constraints.


\end{enumerate}

%
%
%
%
%
%

\section{Background}
\label{why-semantics}


\newcommand{\differentFigPA}{\texttt{debug}}
\newcommand{\differentFigAvOne}{\texttt{4.3.4}}

\newcommand{\differentFigPB}{\texttt{ms}}
\newcommand{\differentFigBvOne}{\texttt{1.0.0}}
\newcommand{\differentFigBvTwo}{\texttt{2.1.0}}
\newcommand{\differentFigBvThree}{\texttt{2.1.2}}

\newcommand{\differentFigConstraintA}{*}
\newcommand{\differentFigConstraintApip}{}
\newcommand{\differentFigConstraintB}{< 2.1.2}


NPM is the most widely used package manager for JavaScript.
(Alternatives such as Yarn are compatible with NPM configurations.)
NPM is co-designed with Node, which is a popular JavaScript runtime for desktop and server applications.
However, NPM is also widely used to manage web applications' dependencies, using ``bundlers'' like Webpack to build programs for the browser.

An NPM configuration (the \texttt{package.json} file) lists package dependencies with version constraints.
NPM has a rich syntax for version constraints that \namenpm{} fully supports.
However, a typical configuration specifies version ranges or exact versions of each dependency.

An unusual feature of NPM is that it may select multiple versions of the same package.
To illustrate, consider the following real-world example that involves the packages \differentFigPA{} and \differentFigPB{} (calendar utilities).
Suppose a project depends on the latest version of \differentFigPB{} (version \texttt{2.1.3}) and \differentFigPA{}.
Unfortunately, \differentFigPA{} depends on an older version of \differentFigPB{} (version \texttt{2.1.2}).
In this situation,  NPM selects both versions of \differentFigPB{}: the root of the project get the latest version, and \differentFigPA{} gets the older version, thus satisfying both constraints independently instead of failing to unify the two constraints. Some package managers handle this situation differently: PIP would report that no solution
exists, while Maven and NuGet would install only the newer version of \texttt{ms} and let \texttt{debug} load
a different version than what it asked for.

Unfortunately, NPM's behavior is not always desirable, and can lead to increased code size
and subtle runtime bugs. Moreover, NPM does not guarantee that packages are
only duplicated when strictly necessary.

Another problem is that what it means to be a \emph{newer package}
is not well-defined across package managers. Suppose package $A$ has versions \texttt{1.0.0} and
\texttt{2.0.0}, and then its author publishes a security numbered \texttt{1.0.1}.
What happens if a program depends on $A$ with no constraints?
Surprisingly, NPM will choose \texttt{1.0.1} because it
was uploaded last, while some other package managers (such as PIP and Cargo) would select \texttt{2.0.0}. Without specific
knowledge of the situation, it is unclear which choice is better.

\subsection{Avoiding Vulnerable Dependencies}
\label{npm-audit-could-be-better}

NPM's built-in tool \texttt{npm audit} checks for vulnerable dependencies by querying the GitHub Advisory Database. The tool can also fix vulnerabilities by upgrading dependencies without violating version constraints.\footnote{The \texttt{--force} flag breaks constraints and potentially breaks the program.} However, the tool has several shortcomings.
\begin{inparaenum}[1)]
  \item Each run only tries to fix a single vulnerability.
  \item It only upgrades vulnerable dependencies, even if a vulnerability-free downgrade is available that respects version constraints.
  \item It does not prioritize fixes by vulnerability scores (CVSS), even though they are available in the GitHub Advisory Database.
  \item It does not make severity-based compromises. For example, a fix may introduce new vulnerabilities that are more severe than the original.
\end{inparaenum}




\subsection{Minimizing Code Bloat}
\label{webpack-could-be-better}

A ``bundler'', such as Webpack, Browserify, or Parcel, is a tool that works in concert with NPM to manage the dependencies of front-end web applications.
The primary task of a bundler is to package all dependencies to be loaded over the web, instead of the local filesystem.
However, bundlers do more, including work to minimize page load times.
A simple way to minimize page load times is to reduce code size.
Unfortunately, NPM's willingness to duplicate packages can lead to increased code size~\cite{inspectpack}.
Contemporary bundlers employ a variety of techniques from minification to unifying individual files with identical contents.
However, these techniques are not always sound and have been known to break widely-used front-end frameworks~\cite{parcel-bug,webpack-side-effects}.

\subsection{Managing Stateful Dependencies}
\label{nobody-understands-peer-dependencies}

NPM's ability to select several versions of the same dependency is also unhelpful when using certain stateful frameworks.
For example, React is a popular web framework that relies on internal global state to schedule view updates.
If a program depends on two packages that transitively depend on two different versions of React, it is likely to encounter runtime errors or silent failures.
The only way to avoid this problem is if all package authors are careful to mark their dependency on React as a \emph{peer dependency}: a dependency that is installed by some other package in a project.
However, there is no easy way to determine that all third-party dependencies use peer dependencies correctly.
It can also be hard to determine before hand that a package will never be used as a dependency and thus should not use peer dependencies.

\section{\namenpm{}}

The goal of \namenpm{} is to help developers address the broad range of problems described above.
\namenpm{} serves as a drop-in replacement for the default \texttt{npm install} command.
The user can run \texttt{npm install --maxnpm} to use \namenpm{}'s customizable dependency solver instead.
There are two broad ways to customize \namenpm{}.
First, \namenpm{} allows the user to specify a prioritized list of objectives with the \texttt{--minimize} flag.
Out of the box, \namenpm{} supports the following objectives (defined precisely in \cref{figure:3-objectives}):
\begin{itemize}
  \item \texttt{min\_oldness}: minimizes the number and severity of installed old versions;
  \item \texttt{min\_num\_deps}: minimizes the number of installed dependencies;
  \item \texttt{min\_duplicates}: minimizes the number of co-installed different versions of the same package; and
  \item \texttt{min\_cve}: minimizes the number and severity of known vulnerabilities.
\end{itemize}
Second, \namenpm{} allows the user to customize how multiple package versions are handled with the \texttt{--consistency} flag:
\begin{itemize}
  \item \texttt{npm}: the default behavior of NPM, which allows several versions of a package to co-exist in a single project; and
  \item \texttt{no-dups}: require every package to have only one version installed.
\end{itemize}
With some work, the user can even
define new objectives and consistency criteria.

For example, a developer building a front-end web application may want to reduce code size and select recent package versions. They could use \namenpm{} as follows:
\begin{minted}[linenos,
  numbersep=5pt,
  fontsize=\footnotesize,
  frame=lines,
  fontfamily=zi4,
  framesep=2mm]{bash}
npm install --maxnpm --consistency no-dups 
  --minimize min_oldness,min_num_deps
\end{minted}
This command avoids duplicating dependencies, and minimizes oldness and the number of dependencies in that order. This is a more principled approach to reducing code size than the ad hoc de-duplication techniques used by bundlers. Moreover, we show that this command is frequently successful at reducing code size (\cref{section:eval:rq_reduce_bloat}).

As a second example, consider a developer building a web application backend, where they are very concerned about security vulnerabilities. They could use \namenpm{} as follows:
\begin{minted}[linenos,
  numbersep=5pt,
  fontsize=\footnotesize,
  frame=lines,
  fontfamily=zi4,
  framesep=2mm]{bash}
npm install --maxnpm --minimize min_cve,min_oldness
\end{minted}
This command subsumes \texttt{npm audit fix} and we show that it is substantially more effective (\cref{section:eval:rq_avoid_vulns}).

\begin{figure}
  \footnotesize
  \[
    \begin{array}{r@{\;}c@{\;}ll}
      \multicolumn{3}{l}{\textbf{Package Metadata}}                                                                                            \\
      \domainPackages     & ::=       & \textbf{\text{String}}                                      & \text{Package Names}                     \\
      \domainVersions     &           & \text{is a set}                                             & \text{Version Numbers}                   \\
      \domainConstraints  &           & \text{is a set}                                             & \text{Version Constraints}               \\


      \domainDependencies & ::=       & \domainPackages \times \domainConstraints                   & \text{Dependencies}                      \\
      \domainNodes'       & \subseteq & \domainPackages \times \domainVersions                      & \text{Package repository (finite set)}   \\
      \domainNodes        & ::=       & \domainNodes' \cup \{ \nodeRoot \}                          & \text{The root node of the solve}        \\

      \funcDeps           & :         & \domainNodes \xrightarrow{\text{fin}} \domainDependencies^* & \text{Dependencies per node}             \\
      \mathcal{M}         & ::=       & \langle \domainNodes, \funcDeps \rangle                     & \text{Package metadata}                  \\[0.5em]

      \multicolumn{3}{l}{\textbf{Solution Graph}}                                                                                              \\
      N_R                 & \subseteq & \domainNodes                                                & \textrm{Package versions in solution}    \\
      \solutionGraphEdges & \in       & \solutionGraphNodes \to \solutionGraphNodes^*               & \text{Solved dependencies}               \\
      \domainGraphs       & ::=       & \langle \solutionGraphNodes , \solutionGraphEdges \rangle   & \text{Solution graphs}                   \\[0.5em]

      \multicolumn{4}{l}{\textbf{Dependency Solver Specification}}                                                                             \\
      \funcSat            & :         & \domainConstraints \to \domainVersions \to \domainBools     & \text{Constraint satisfaction semantics} \\
      \funcConsistent     & :         & \domainVersions \to \domainVersions \to \domainBools        & \text{Version consistency versions}      \\
      \funcMin            & :         & \domainGraphs \to \domainReals^n                            & \text{Objective functions}
    \end{array}
  \]

  \caption{The \namebackend Model of Dependency Solving}
  \label{math:pacsolve-abstract-io}
\end{figure}

\label{pacsolve}

\namenpm{} is built on \namebackend{}, which is a DSL for describing dependency solvers.
This section presents \namebackend{} with an emphasis on how we use it to build \namenpm{}. 
\cref{discussion} discusses the potential applicability of \namebackend{} to building solvers for other languages.

\subsection{Describing a Dependency Solver with \namebackend}

\namebackend{} is a solver-aided language built with Rosette, a solver-aided programming language that interfaces with the Z3 theorem prover~\cite{torlak2013growing}.
The input to \namebackend{} is a set of available packages with their dependencies. The output is a \emph{solution graph} where nodes are specific versions of a package and edges represent dependencies. (\Cref{semantics} formally describes the semantics and relationship between available packages and solution graphs.)
Without loss of generality, we assume there is a distinguished root package ($\nodeRoot$).

To build a dependency solver with \namebackend{}, we have to define:
\begin{enumerate}

  \item The abstract syntax of package versions ($\domainVersions$) and version constraints ($\domainConstraints$), which tends to vary subtly between dependency solvers;

  \item The \emph{constraint satisfaction predicate} that determines if a given version satisfies a constraint ($\funcSat$);

  \item The \emph{version consistency predicate} that consumes two package
        versions and determines if those two versions of the same package may be co-installed
        ($\funcConsistent$); and

  \item The \emph{objective function} that determines the cost of a solution graph ($\funcMin$).

\end{enumerate}

\namebackend{} produces a solution that is optimal and consistent with all constraints.
A solution graph may optionally have cycles, which some package managers allow (including NPM).

\begin{figure}
  \centering

  \begin{subfigure}{\columnwidth}
    \footnotesize
    \(
    \begin{array}{r@{\;}c@{\;}ll}
      \domainVersions    & \pdef & (x\ y\ z)                                                  & \text{Version numbers}        \\
      \domainConstraints & \pdef & (\texttt{=}\ x\ y\ z)                                      & \text{Exact}                  \\
                         & \mid  & \texttt{*}                                                 & \text{Any}                    \\
                         & \mid  & (\texttt{<=}\ x\ y\ z)                                     & \text{At most}                \\
                         & \mid  & (\texttt{>=}\ x\ y\ z)                                     & \text{At least}               \\
                         & \mid  & (\wedge\ x\ y\ z)                                          & \text{Semver compatible with} \\
                         & \mid  & (\texttt{and}\ \domainConstraints_1\ \domainConstraints_2) & \text{Conjunction}            \\
                         & \mid  & (\texttt{or}\ \domainConstraints_1\ \domainConstraints_2)  & \text{Disjunction}            \\
    \end{array}
    \)
    \caption{Example of $\domainVersions$ and $\domainConstraints$, allowing conjunction, disjunction, and range operators on semver-style versions.}
    \label{listing:funcs-example-domains}
  \end{subfigure}

  \vskip 1em
  \hrule
  \vskip 1em

  \begin{subfigure}{0.9\columnwidth}
    \begin{minted}[
  linenos,
  numbersep=5pt,
  fontfamily=zi4,
  fontsize=\footnotesize]{racket}
(define (sat c v)
  (match `(,v ,c)
    [`((,x ,y ,z) (= ,x ,y ,z))        #true]
    [`(,_ *)                           #true]
    [`((,x ,y ,z1) (<= ,x ,y ,z2))     (<= z1 z2)]
    [`((,x ,y1 ,z1) (<= ,x ,y2 ,z2))   (< y1 y2)]
    [`((,x1 ,y1 ,z1) (<= ,x2 ,y2 ,z2)) (< x1 x2)]
    [`((0 0 ,z1) (^ 0 0 ,z2))          (= z1 z2)]
    [`((0 ,y ,z1) (^ 0 ,y ,z2))        (>= z1 z2)]
    [`((0 ,y1 ,z1) (^ 0 ,y2 ,z2))      #false]
    [`((,x ,y ,z1) (^ ,x ,y ,z2))      (>= z1 z2)]
    [`((,x ,y1 ,z1) (^ ,x ,y2 ,z2))    (> y1 y2)]
    [`(,_ (and ,c1 ,c2))
      (and (sat c1 v) (sat c2 v))]
    [`(,_ (or ,c1 ,c2))
      (or (sat c1 v) (sat c2 v))]
    [`((,x1 ,y1, z1) (>= ,x2 ,y2 ,z2))
      (sat `(,x2 ,y2 ,z2) `(<= ,x1 ,y1 ,z1))]
    [_                                 #false]))

(define (consistent v1 v2)
  #true)
\end{minted}
    \caption{A constraint satisfaction predicate and a version consistency predicate for $\domainVersions$ and $\domainConstraints$ defined in \cref{listing:funcs-example-domains}.}
    \label{listing:funcs-example-sat}
  \end{subfigure}

  \caption{The syntax of versions and constraints, and the constraint satisfaction predicate and consistency predicate for a fragment of NPM.}
  \label{listing:funcs-example}
\end{figure}

\begin{figure*}
  \centering
  \begin{minipage}{0.48\textwidth}
    \begin{subfigure}{\textwidth}
      \begin{minted}[
  linenos,
  numbersep=5pt,
  fontsize=\footnotesize,
  fontfamily=zi4,
  framesep=2mm]{racket}
(define (minGoal-num-deps g)
  (length (graph-nodes g)))
\end{minted}
      \caption{\footnotesize Minimize the total number of installed dependencies.}
      \label{listing:objective-min-deps}
    \end{subfigure}

    \begin{subfigure}{\textwidth}
      \phantom{.} 
      \vspace{10pt}

      \begin{minted}[
  linenos,
  numbersep=5pt,
  fontsize=\footnotesize,
  fontfamily=zi4,
  framesep=2mm]{racket}
(define (minGoal-duplicates g)
  ; we count how many times
  ; each package name occurs
  (define package-counts (foldl
    (lambda (n counts)
      (define p (node-package n))
      (hash-set counts p
        (add1 (hash-ref counts p 0))))
    (make-immutable-hash)
    (graph-nodes g)))

  ; then assign a cost of 1
  ; for each duplicate
  (apply +
    (map
      (lambda (c) (max 0 (sub1 c)))
      (hash-values package-counts))))
\end{minted}
      \caption{
        \footnotesize Minimize the total number of
        co-installed versions of the same package.
      }
      \label{listing:objective-min-duplicates}
    \end{subfigure}
  \end{minipage}
  \quad\vline\quad\quad
  \begin{minipage}{0.43\textwidth}
    \begin{subfigure}{\textwidth}
      \begin{minted}[
  linenos,
  numbersep=5pt,
  fontsize=\footnotesize,
  fontfamily=zi4,
  framesep=2mm]{racket}
(define (minGoal-oldness g)
  (apply +
    (map
      (lambda (n)
        (get-oldness
          (node-package n)
          (node-version n)))
      (graph-nodes g))))

(define (get-oldness p v)
  ; The get-sorted-versions retrieves
  ; a list of all versions of p
  (define all-vs
    (get-sorted-versions p))
  (if (= (length all-vs) 1)
      0
      (/ (index-of all-vs v)
          (sub1 (length all-vs)))))
\end{minted}
      \caption{
        \footnotesize Minimize the amount of
        ``oldness'' present in the solution graph.
        Each resolved dependency contributes an oldness proportional
        to its rank among the total ordering of versions of that package
      }
      \label{listing:objective-min-oldness}
    \end{subfigure}
  \end{minipage}

  \caption{Three different examples of \namebackend minimization objectives}
  \label{figure:3-objectives}
\end{figure*}
\paragraph*{A Fragment of NPM}

\Cref{listing:funcs-example-domains} shows an example of versions and constraints for a fragment of NPM.
(\namenpm{} supports the full syntax and takes care of parsing NPM's concrete syntax to the parenthesized syntax that Rosette and \namebackend{} require.)
Given this syntax of versions and constraints, \cref{listing:funcs-example-sat} shows constraint satisfaction and version consistency predicates. The \texttt{sat} function receives as input a version constraint and a version
(from the syntax of \Cref{listing:funcs-example-domains}) and returns a boolean by performing a \texttt{match}
case analysis.
Interesting subtleties include lines 8--12 handling the matching semantics of caret constraints (\texttt{\^{}1.2.3}) including their different behavior for versions with leading zeros, and lines 13--16 performing structural recursion
on the constraints.
The \texttt{consistent} function receives two versions and must determine if they can be co-installed.
This simple consistency predicate is the constant true function, which implements NPM's standard policy of always
allowing co-installations.

\Cref{figure:3-objectives} defines three examples of objective functions that consume a solution graph and produce a cost. \namebackend{} minimizes cost, so
a trivial objective is to minimize the total number of packages, completely ignoring package age and other factors (\cref{listing:objective-min-deps}).
An alternative objective is to minimize the co-installation of multiple versions of the same package (\cref{listing:objective-min-duplicates}).
The function counts the number of versions of each package, and assigns a cost to every package that has more than one version.
Our final example is an interpretation of the common goal that package managers have of trying to choose newer versions of dependencies (\cref{listing:objective-min-oldness}).
The function gives each node an \emph{oldness score} between $0$ (newest) and $1$ (oldest), with the scores evenly divided across all versions of a package in the solution.
There are two subtleties with this definition.
1)~We perform \emph{minimization}, since maximization would encourage the solver to find large solutions that inflate newness.
2)~We take the sum rather than the mean, since taking the mean would also encourage the solver to add extra packages that deflate oldness.

Other metrics are possible as well, such as our implementation of an aggregated score of dependency vulnerabilities (\cref{section:eval:rq_avoid_vulns}), or the total download size~\cite{tucker+:icse07-opium}.
With a library of optimization objective functions defined, \namebackend{} then allows for easy composition
of objective functions either by multiple prioritized objectives, or by weighted linear combinations of objectives.

\subsection{The Semantics of \namebackend}
\label{semantics}

This section formally describes the semantics of \namebackend{}.
The \emph{package metadata} (top of
\cref{math:pacsolve-abstract-io}) is described by 1)~a set of package name
and version pairs ($\domainNodes$), and 2)~a map ($\funcDeps$) from these pairs to a list of dependencies.
Each dependency
specifies a package name and version constraint ($\domainConstraints$).

The solution graph is a directed graph where the
nodes are package-version pairs, and each node has an ordered list of edges.
The order of edges corresponds to the order of dependencies in
the package metadata.

The semantics of \namebackend{} is a relation ($\mathcal{S}$) between the package metadata, the dependency solver specification, and the solution graph.
The relation holds
when a solution graph is valid with respect to the package metadata
and the dependency solver specification. The relation holds if and only if
the following six conditions are satisfied. First, the solution graph must
include the root:
\begin{equation}
  \nodeRoot \in \solutionGraphNodes
\end{equation}
Second, the solution graph must be connected, to ensure it does not
have extraneous packages:
\begin{equation}
  \langle N_R, D_R \rangle~\textrm{is connected}
\end{equation}
Third, for all packages in the solution graph, every edge must correspond
to a constraint in the package metadata:
\begin{equation}
  \forall n. n \in N_R \implies |\solutionGraphEdges(n)| = |\funcDeps(n)|
\end{equation}
Fourth, for every edge in the solution graph that points to package $p$ with
version $v$, the corresponding constraint in the package metadata must
refer to package $p$ with constraint $c$, where $v$ satisfies $c$:
\begin{equation}
  \begin{array}{r@{\,}l}
    \forall n .     & n \in N_R \implies                                    \\
                    & \forall i. 0 \le i < |D_R(n)| \implies                \\
    \exists p,v,c . & \; (p, v) = \solutionGraphEdges(n)[i]                 \\
                    & \wedge (p, c) = \funcDeps(n)[i] \wedge \funcSat(c, v)
  \end{array}
\end{equation}

The criteria so far are adequate for many dependency solvers, but permits solutions
that may be unacceptable. For example, without further constraints, a solution
graph may have several versions of the same package. Thus the fifth condition
ensures that if there are versions of a package in the solution graph, then
the two versions are consistent, as judged by the dependency solver specification:
\begin{equation}
  \forall p,v,v'. \; (p, v), (p, v') \in \solutionGraphNodes \implies \funcConsistent(v, v')
\end{equation}
NPM allows arbitrary versions to be co-installed (so $\funcConsistent$ is the constant true function),
PIP only allows exactly one version of a package to be installed at
a time (so $\funcConsistent$ requires $v = v'$), and
Cargo only allows semver-incompatible versions to be co-installed.

A final distinction between dependency solvers is whether or not they
allow cyclic dependencies. NPM and PIP allow cycles, but others, such as Cargo, do not.
Thus the sixth condition, which is \emph{optional}, uses the dependency solver specification to determine
whether or not cycles are permitted:
\begin{equation}
  \langle N_R, D_R \rangle~\textrm{is acyclic}
\end{equation}

These five or six conditions determine whether or not a solution graph is correct
with respect to the semantics of a particular dependency solver.

\subsection{Synthesizing Solution Graphs with \namebackend{}}
\label{sol-graph-synth}
\lstset{language=Racket}

Dependency solving with possible conflicts is NP-complete~\cite{dicosmo:edos}.
Some package managers use polynomial-time algorithms by giving up
on various properties, such as disregarding conflicts (NPM)
and eschewing completeness (NPM and PIP's old solver~\cite{pip-new-resolver}).
Since the \namebackend{} model includes a generalized notion of
conflicts ($\funcConsistent$), we leverage Max-SMT solvers to
implement \namebackend{} effectively.

We implement \namebackend{} in Rosette, which is a solver-aided
language that facilitates building verification and synthesis tools for DSLs.
In the \namebackend{} DSL, \emph{the program is a solution graph}. We implement
a function that consumes
\begin{inparaenum}
  \item package metadata,
  \item a dependency solver specification (\cref{math:pacsolve-abstract-io}) and
  \item a solution graph,
\end{inparaenum}
and then asserts that the solution graph is correct.
With a little effort, we can replace the input solution graph with a \emph{solution graph sketch}. This allows us to use Rosette to perform
\emph{angelic execution}~\cite{broy1981algebraic}
to synthesize a solution graph that satisfies the correctness criteria.
This section describes the synthesis
procedure in more detail, starting with how we build a sketch.

\paragraph{Sketching solution graphs}

Before invoking the Rosette solver, we build a sketch of a solution graph
that has a node for every version of every package that
is reachable from the set of root dependencies.
Every node in a sketch has the following fields:
\begin{inparaenum}
  \item a concrete name and version for the package that it represents;
  \item a symbolic boolean \emph{included} that indicates whether or not the
  node is included in the solution graph;
  \item a symbolic natural number \emph{depth} which we use to enforce acyclic
  solutions when desired;
  \item a vector of concrete dependency package names;
  \item a vector of concrete version constraints for each dependency; and
  \item a vector of symbolic \emph{resolved versions} for each dependency.
\end{inparaenum}

\cref{figure:graph-sketch} illustrates an example solution graph sketch
corresponding to a dependency solving problem involving two packages (\texttt{debug}, \texttt{ms}),
where \texttt{ms} is a dependency of both \texttt{debug} and the root, while \texttt{debug} is depended on
by only the root.
The combination of concrete dependency names and symbolic dependency
versions can be seen as representing symbolic edges in two parts:
a concrete part which does not need to be solved for (solid arrows),
and a symbolic part which requires solving (dashed arrows). This representation
shrinks the solution space of graphs as outgoing edges
can only point to nodes with the correct package name.

\pgfdeclarelayer{background}
\pgfdeclarelayer{foreground}
\pgfsetlayers{background,main,foreground}


\tikzstyle{node}=[draw, fill=blue!20, text width=6em,
     minimum height=2.5em,drop shadow]
\tikzstyle{attribute}=[draw, fill=green!20, 
    text centered, 
    drop shadow, rounded corners]
\tikzstyle{root}=[draw, fill=red!20, text width=5em,
    text centered, minimum height=2.5em,drop shadow,rounded corners]

\tikzstyle{concreteEdge} = [draw, -{Latex[length=3mm,width=3mm]}, line width=0.3mm]
\tikzstyle{symbolicEdge} = [draw, -{Latex[length=2mm,width=2mm]}, line width=0.2mm, dashed]

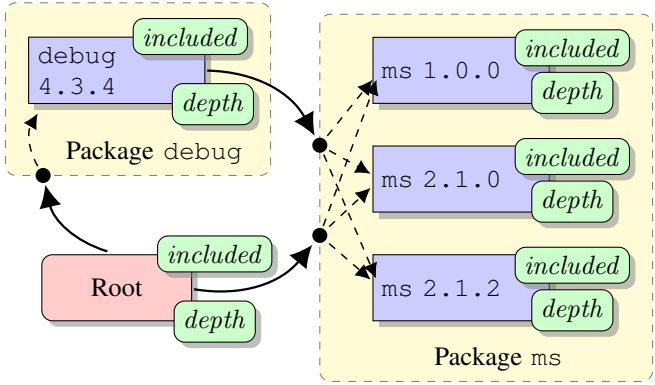
\begin{figure}
  \centering
  \begin{tikzpicture}
    \node (a101) [node] {\texttt{\differentFigPA{} \differentFigAvOne{}}};
    \path (a101.east)+(-0.2,+0.45) node (a101inc) [attribute] {$\mathit{included}$};
    \path (a101.east)+(0.1,-0.45) node (a101depth) [attribute] {$\mathit{depth}$};
    \path (a101.south west)+(0.2,0.0) node (a101Port) {};
    \path (a101.east)+(-0.1, 0) node (a101Out) {};

    \begin{pgfonlayer}{background}
      \path (a101.west |- a101.north)+(-0.3,0.45) node (a) {};
      \path (a101.west |- a101.south)+(-0.3,-0.95) node (b) {};
      \path (a101.south -| a101inc.east)+(+0.3,-0.95) node (c) {};

      \path[fill=yellow!20,rounded corners, draw=black!50, dashed]
          (a) rectangle (c);
      \path (a101.south)+(0.5,-0.65) node {Package \differentFigPA};
    \end{pgfonlayer}

    \path (b)+(0.5, 0) node (aPort) {};
    \fill (aPort) circle (0.1cm);

    \path (a101.east)+(3.4,0) node (b101) [node] {\texttt{\differentFigPB{} \differentFigBvOne{}}};
    \path (b101.east)+(0.3,+0.3) node (b101inc) [attribute] {$\mathit{included}$};
    \path (b101.east)+(0.3,-0.3) node (b101depth) [attribute] {$\mathit{depth}$};
    \path (b101.west)+(0.1,0) node (b101Port) {};

    \path (b101.south)+(0.0,-1.0) node (b201) [node] {\texttt{\differentFigPB{} \differentFigBvTwo{}}};
    \path (b201.east)+(0.3,+0.3) node (b201inc) [attribute] {$\mathit{included}$};
    \path (b201.east)+(0.3,-0.3) node (b201depth) [attribute] {$\mathit{depth}$};
    \path (b201.west)+(0.1,0) node (b201Port) {};

    \path (b201.south)+(0.0,-1.0) node (b202) [node] {\texttt{\differentFigPB{} \differentFigBvThree{}}};
    \path (b202.east)+(0.3,+0.3) node (b202inc) [attribute] {$\mathit{included}$};
    \path (b202.east)+(0.3,-0.3) node (b202depth) [attribute] {$\mathit{depth}$};
    \path (b202.west)+(0.1,0) node (b202Port) {};

    \begin{pgfonlayer}{background}
      \path (b101.west |- b101.north)+(-0.7,0.3) node (a) {};
      \path (b101.west |- b202.south)+(-0.7,-0.8) node (b) {};
      \path (b202.south -| b202inc.east)+(+0.3,-0.8) node (c) {};

      \path[fill=yellow!20,rounded corners, draw=black!50, dashed]
          (a) rectangle (c);
      \path (b202.south)+(0.5,-0.5) node {Package \differentFigPB{}};
    \end{pgfonlayer}

    \path (a) -- node (bPortMid) {} (b);
    \path (bPortMid)+(0,-0.5) node (bPortRoot) {};
    \fill (bPortRoot) circle (0.1cm);
    \path (bPortMid)+(0,0.7) node (bPortA) {};
    \fill (bPortA) circle (0.1cm);

    \path (a101 |- b202) node (root) [root] {Root};
    \path (root.east)+(0.3,+0.45) node (rootinc) [attribute] {$\mathit{included}$};
    \path (root.east)+(0.3,-0.45) node (rootdepth) [attribute] {$\mathit{depth}$};

    \path (root.north)+(0,0) node (rootOut1) {};
    \path (root.east)+(-0.1,0) node (rootOut2) {};


    \path [concreteEdge] (rootOut1) edge[bend left] node []{} (aPort);
    \path [concreteEdge] (rootOut2) edge[bend right] node []{} (bPortRoot);
    \path [symbolicEdge] (aPort) edge[bend left] node []{} (a101Port);

    \path [concreteEdge] (a101Out) edge[bend left] node []{} (bPortA);

    \path [symbolicEdge] (bPortRoot) edge[] node []{} (b101Port);
    \path [symbolicEdge] (bPortRoot) edge[] node []{} (b201Port);
    \path [symbolicEdge] (bPortRoot) edge[] node []{} (b202Port);

    \path [symbolicEdge] (bPortA) edge[] node []{} (b101Port);
    \path [symbolicEdge] (bPortA) edge[] node []{} (b201Port);
    \path [symbolicEdge] (bPortA) edge[] node []{} (b202Port);
  \end{tikzpicture}
  \caption{A sample solution graph sketch}
  \label{figure:graph-sketch}
\end{figure}

\paragraph{Graph Sketch Solving}
We define three assertion functions that check
correctness criteria of a solution graph:
\begin{enumerate}

  \item \lstinline|check-dependencies| 
        asserts that if a node is \emph{included}, then
        all the dependencies of the node are
        \emph{included} and
        satisfy the associated
        version constraints as judged by the
        constraint interpretation function ($\funcSat$).
        We run this assertion function on all nodes,
        and additionally assert that the root node is \emph{included}.

  \item \lstinline|globally-consistent?| asserts that
        the consistency function ($\funcConsistent$) holds on all pairs of nodes with the same package name.

  \item \lstinline|acyclic?| asserts
        that the \emph{depth} of a node is strictly less than the depth of all
        its dependencies.
        If an acyclic solution is desired, we run this assertion function
        on all nodes, and assert that the root node has depth
        zero.
\end{enumerate}
We then ask Rosette to find a concrete solution graph that satisfies the above constraints
while minimizing the objective function ($\funcMin$).
As a final step, we traverse the concretized solution graph sketch
from the root node, and install on-disk
those nodes that are marked \emph{included}.

\section{Evaluation}
\label{section:evaluation}

We evaluate \namenpm in several scenarios, determining whether its support for flexible optimization objectives can provide tangible benefits to developers as compared to NPM.
We gather two large datasets of popular packages, and also investigate if \namenpm is sufficiently reliable and performant to use as a drop-in replacement for NPM.
In this section we use \namenpm{} to answer each of our research questions:
\begin{description}
  \setlength{\itemindent}{-.5in}
  \item \rqOne
  \item \rqTwo
  \item \rqThree
  \item \rqFour
\end{description}

We build two datasets of NPM packages.
\datasetTopName{} is a set of the latest versions of the top 1,000 most-downloaded packages as of August 2021.
Including their dependencies, there are 1,147 packages in this
set.
Unsurprisingly, these packages are maintained and have few known vulnerabilities.
Therefore, to evaluate vulnerability mitigation, we build the \datasetVulnName{} dataset of 715 packages with high CVSS scores as follows:
\begin{inparaenum}
  \item we filter the \datasetTopName{} to only include packages with available GitHub repositories;
  \item we extract every revision of \texttt{package.json};
  \item for each revision, we calculate the aggregate CVSS score of their direct dependencies, as determined by the GitHub Advisory Database; and
  \item we select the highest scoring revision of each package.
\end{inparaenum}

\namenpm{} is built on NPM 7.20.1. \namebackend{} uses Racket 8.2, Rosette commit \texttt{1d042d1}, and Z3 commit \texttt{05ec77c}.
We configure
NPM to not run post-install scripts and not install optional dependencies.
We run our performance benchmarks on Linux, with a 16-Core AMD EPYC 7282 CPU with 64 GB RAM. We warm the NPM local package cache before measuring running times.

\subsection{\rqone}

\subsubsection{\rqOneDotOne}
\label{section:eval:rq_avoid_vulns}

\begin{figure*}
\begin{center}

\begin{subfigure}{0.45\textwidth}
    \includegraphics[width=\columnwidth]{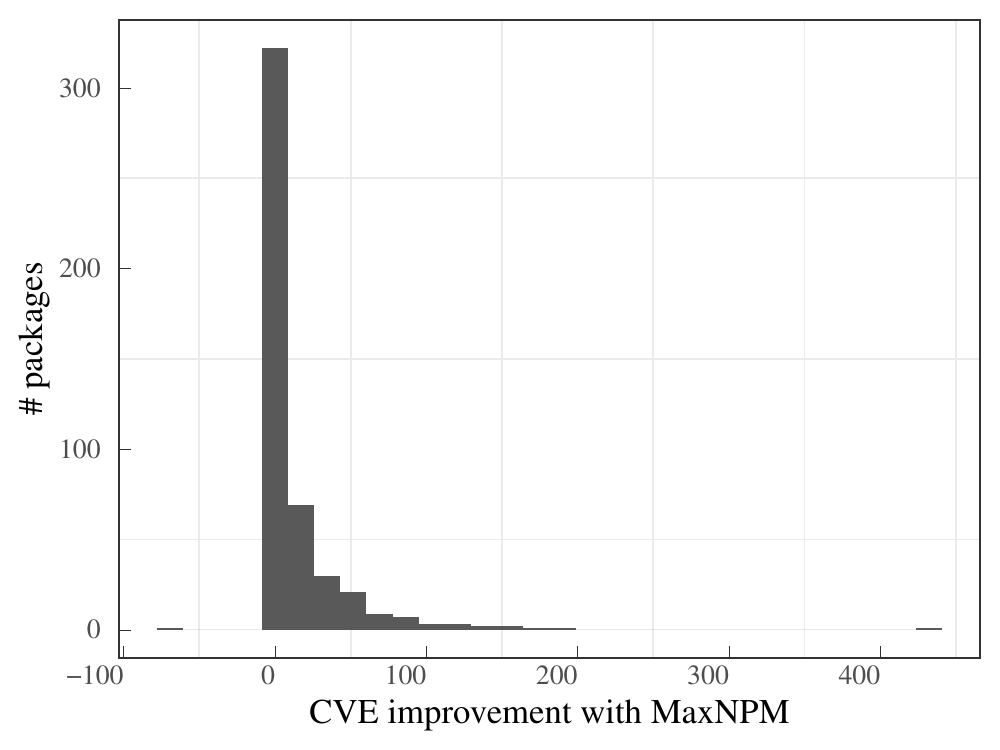}
    \caption{
        A histogram showing CVSS improvement of packages solved with \namenpm{} (configured to minimize vulnerabilities first) 
        compared to NPM's auditing tool. CVSS decreases with \dataCveExpMaxNpmBetterTotalPerc{} of packages.
    }
    \label{figure:cves_hist}
\end{subfigure}
\quad\quad
\begin{subfigure}{0.45\textwidth}
\includegraphics[width=\columnwidth]{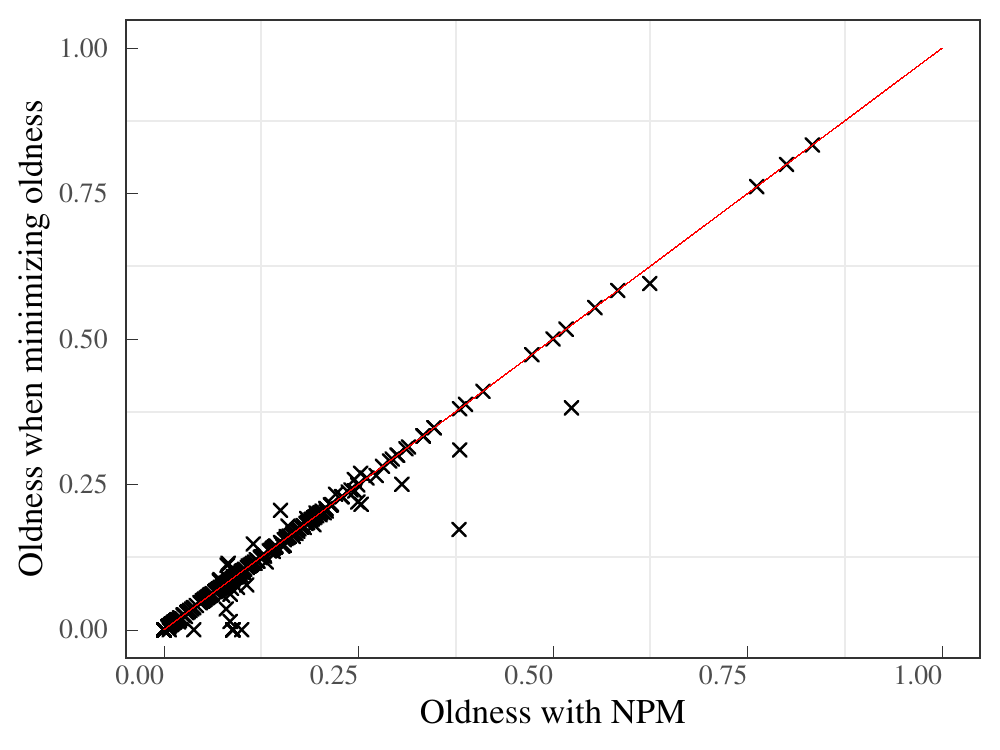}
\caption{
    \namenpm is configured to minimize
    oldness and the number of dependencies (in that order).
    Each point represents a package, and those below the line
    have newer dependencies, by the metric in \Cref{section:evaluation-better-solutions}. 
    Overall, \namenpm{} finds newer versions for dependencies.
}
\label{figure:oldness}
\end{subfigure}

\begin{subfigure}{0.45\textwidth}
\includegraphics[width=\columnwidth]{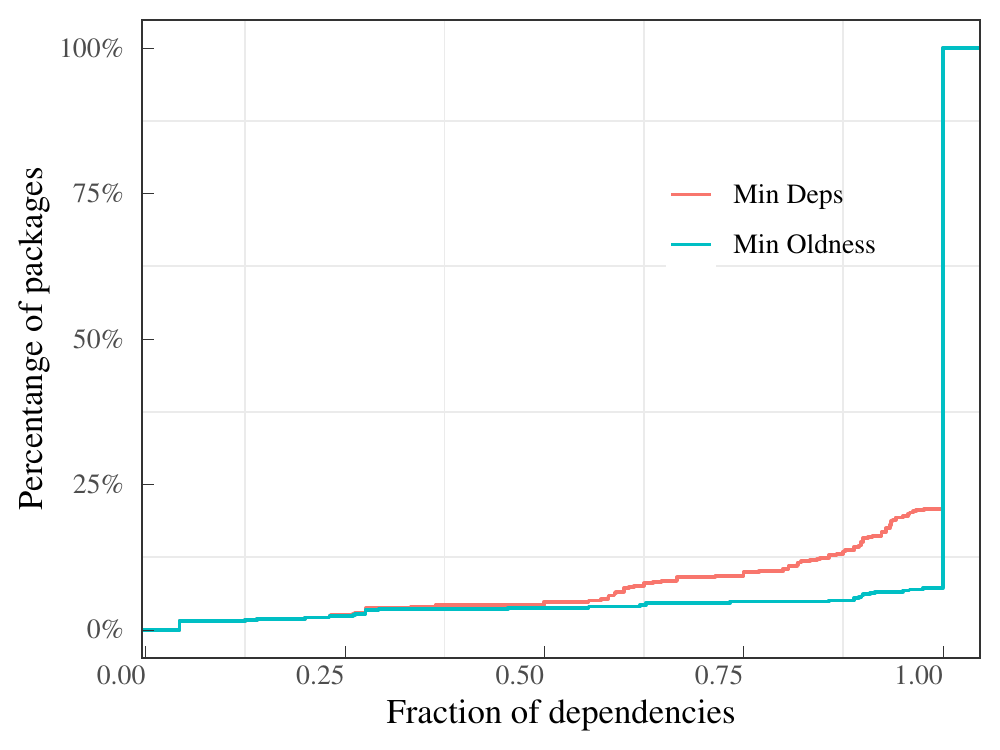}
\caption{
    An ECDF of the fraction of dependencies compared to NPM, when \namenpm{}
    is configured to
    \begin{inparaenum}[a)]
        \item minimize dependencies then oldness (red ECDF), or
        \item minimize oldness then dependencies (blue ECDF).
    \end{inparaenum}
    \namenpm can reduce dependencies
    in about \dataFractionShrinking{} of packages.
}
\label{figure:shrinkage}
\end{subfigure}
\quad\quad
\begin{subfigure}{0.45\textwidth}
\includegraphics[width=\columnwidth]{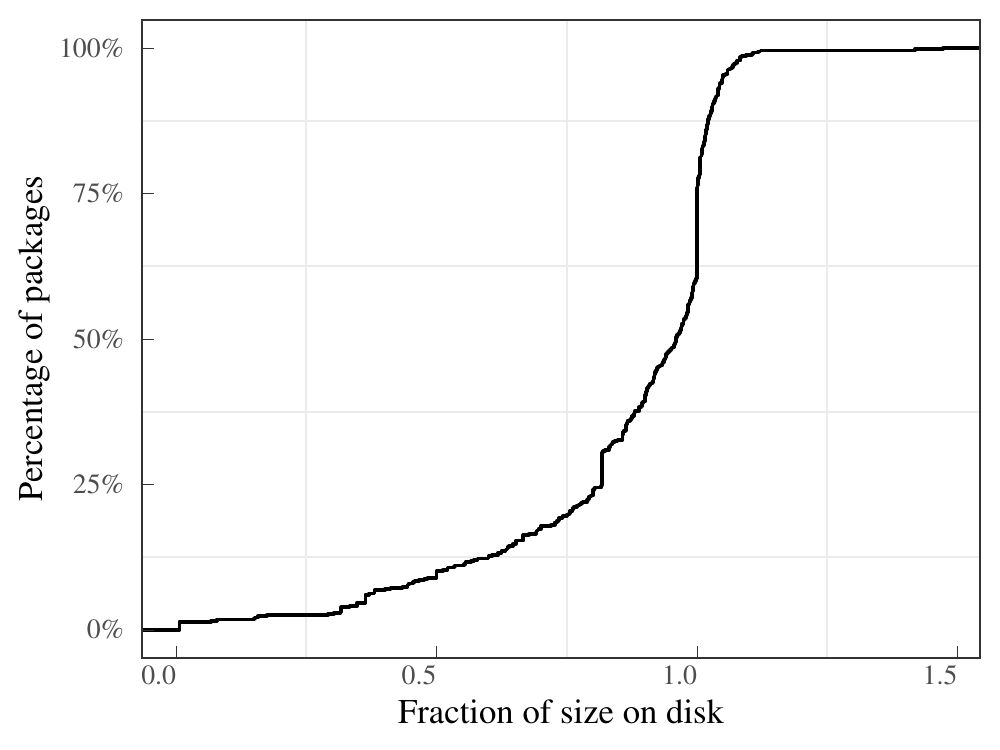}
\caption{
    An ECDF of the ratio of disk space of packages
    solved using \namenpm{} (configured to minimize number of dependencies, then oldness) 
    vs NPM. \namenpm{} can reduce space required.
}
\label{disk_shrinkage}
\end{subfigure}
\end{center}

\caption{
    Comparing NPM's to \namenpm{}'s solution quality. 
    These plots ignore failures in both solvers
    and have \namenpm{} configured to use NPM-style consistency
    and allow cycles.
}
\end{figure*}

We configure \namenpm to minimize the aggregate CVSS scores of all dependencies,\footnote{The \texttt{min\_cve,min\_oldness} flags.} and compare with the built-in \texttt{npm audit fix} tool (\cref{npm-audit-could-be-better}).
We use the \datasetVulnName{} packages for this comparison.
Both tools run successfully on \dataCveExpSuccessCount{} packages: the failures occur because these are typically older versions that do not successfully install.

The histogram in \cref{figure:cves_hist} reports the difference in aggregate CVSS score between \texttt{npm audit fix} and \namenpm. 
A higher score indicates that a package has fewer vulnerabilities with \namenpm.
\namenpm produces fewer vulnerabilities on \dataCveExpMaxNpmBetterCount{} packages (\dataCveExpMaxNpmBetterTotalPerc{}).
There is one package where \namenpm produces a lower score; we are investigating this as a possible bug.
The mean CVSS improvement by \namenpm{} is \dataCveExpMeanCveDiff{}
CVSS points (a ``maximum severity'' vulnerability is 10 points), or by 30.51\%.
The improvement is statistically significant ($p<2.2 \times 10^{-16}$) using a paired Wilcoxon signed rank test, with a medium Cohen's d effect size of $d=0.46$.
Thus, we find that \namenpm{} is substantially more effective than \texttt{npm audit fix} at removing vulnerable dependencies.

An example project where \namenpm{} eliminates vulnerabilities is the \texttt{babel} compiler (34 million weekly downloads). Commit \texttt{5b09114b8} is in \datasetVulnName, and \namenpm{} eliminates all vulnerabilities; whereas \texttt{npm audit fix} leaves several with an aggregate CVSS score of 59.4.

\subsubsection{\rqOneDotTwo}
\label{section:evaluation-better-solutions}

\namenpm{} ought to be able to find newer packages than NPM's greedy algorithm.
We define the oldness of a dependency on a package version ($\mathit{old}(p, v)$) as a
function that assigns the newest version the value $0$, the oldest version the
value $1$, and other versions on a linear scale in between. We define the mean oldness of project as the mean oldness of all dependencies in a project,including transitive dependencies.
Note that this metric is not identical to the minimization objective of \namenpm, which calculates the sum and ignores duplicates. The metric is more natural to interpret, whereas the objective function avoids pathological solutions.

\Cref{figure:oldness} shows a point for every package in the \datasetTopName{}, with mean oldness using
NPM and \namenpm as its $x$ and $y$ coordinates. Points on
$y=x$ are packages whose dependencies are just as old with both NPM and
\namenpm. 
\dataFractionNewer{} of packages excluding those with zero dependencies are better with \namenpm{}, while
\dataFractionOlder{} are worse. On average oldness improved by 2.62\%.
The improvement is statistically significant ($p=4.27 \times 10^{-6}$) using a paired Wilcoxon signed rank test, with a small Cohen's d effect size of $d=0.024$.
Thus \namenpm produces newer dependencies on average.

An example of successful oldness minimization is the
\texttt{class-utils} package (15 million weekly downloads).
\namenpm{} chooses a slightly older version of a direct dependency,
which allows it to chose much newer versions of transitive dependencies.

One might wonder why \namenpm{} does worse in \dataFractionOlder{} of cases, since \namenpm{} should be optimal.
Manual investigation of these cases shows that some packages make use of features which we have not implemented in \namenpm{},
such as URLs to tarballs rather named dependencies. \namenpm{} is unable to explore that
region of the search space. Implementing these features would take some engineering effort, but wouldn't require changes to the model.

\subsubsection{\rqOneDotThree}
\label{section:eval:rq_reduce_bloat}

Instead of using ad hoc and potentially unsound techniques to reduce code bloat, we can configure \namenpm{} to minimize the total number of dependencies.
On the \datasetTopName{} packages, we configure \namenpm{} in two ways:
1)~prioritize fewer dependencies over lower oldness; and 2)~prioritize lower oldness over fewer dependencies.
\Cref{figure:shrinkage} plots an ECDF (empirical cumulative distribution function) plot where the $x$-axis shows
the shrinkage in dependencies compared to NPM, and the $y$-axis shows the cumulative percentage of packages with that amount of shrinkage. 
The plot excludes packages with zero dependencies, and $x=1$ indicates no shrinkage (when \namenpm{} produces just as many dependencies as NPM).

Both configurations produce fewer dependencies than NPM, but prioritizing fewer dependencies is the most effective (the red line).
For about \dataFractionShrinking{} of packages,
\namenpm is able to reduce the number of dependencies, with an average reduction in the number of packages of 4.37\%.
The improvement is statistically significant ($p<2.2 \times 10^{-16}$) using a paired Wilcoxon signed rank test, with a moderate Cohen's d effect size of $d=0.20$.
For the same set of packages, the total disk space required
shrinks significantly (\cref{disk_shrinkage}). With \namenpm{}, a quarter of the packages require
\dataFSShrinkageQuartileFirst{} of their original disk space.
Even when we prioritize lowering oldness, \namenpm{} still produces fewer dependencies (the blue line).


An example of dependency size minimization is the
\texttt{assert} package (13 million weekly downloads).
For 3 direct dependencies, \namenpm{} chooses slightly older revisions (with the same major and minor version). But this eliminates 33 of 43 transitive dependencies.

We have observed that in a few cases NPM exhibits a bug 
in which it installs additional dependencies 
that are not defined in the set of production dependencies of the package,
nor in the set defined by transitive dependencies.
\footnote{\texttt{@babel/plugin-proposal-export-namespace-from} is an example.}
However, \namenpm{} does not exhibit this bug.
We will work on reporting this bug, but we believe this an example of the advantage of 
\namebackend{}'s declarative style of building package managers.

\subsubsection{\rqOneDotFour}
\label{section:evaluation:tree_solving}

\begin{table*}[htbp]
  \begin{center}
    \begin{tabular}{lllllrrrr}
      \toprule 
        &  &   & \multicolumn{2}{c}{Minimization Objectives}  &  & \multicolumn{3}{c}{Failures} \\       \cmidrule{4-5} \cmidrule{7-9} 
        Solver & Consistency & Allow cycles? & Primary & Secondary &  Successes& Unsat & Timeout & Other \\ 
      \midrule
        NPM &  &  &  & & 953 &   0 &   0 &  47 \\ 
        \namenpm{} & npm & Yes & Oldness& \# of Dependencies & 972 &   0 &  27 &   1 \\ 
        \namenpm{} & npm & Yes & \# of Dependencies & Oldness & 972 &   0 &  27 &   1 \\


        \namenpm{} & npm & Yes & Oldness & Duplicate Packages & 973 &   0 &  26 &   1 \\ 


        \namenpm{} & no-dups & Yes & Oldness & \# of Dependencies & 926 &  19 &  54 &   1 \\ 




        \namenpm{} & npm & No & Oldness & \# of Dependencies & 972 &   0 &  27 &   1 \\ 
        \namenpm{} & no-dups & No & Oldness & \# of Dependencies & 926 &  19 &  54 &   1 \\ 
      \bottomrule
    \end{tabular}
  \end{center}
  \caption{Failures that occur when running NPM and different configurations of \namenpm on the \datasetTopName{} dataset.}
  \label{table:failures}
\end{table*}

NPM happily allows a program to load several versions of the same package, which can lead to subtle bugs (\cref{nobody-understands-peer-dependencies}).
To address this problem, a developer can configure \namenpm{} to disallow duplicates.
In this configuration, \dataNumPIPUnsupported{} packages (\dataFractionPIPUnsupported{}) in \datasetTopName{} produce unsatisfiable constraints, which indicates that they require several versions of some package (\Cref{table:failures}).

For example, \texttt{terser@5.9.0} is a widely used JavaScript parser that that directly depends on \texttt{source-map@0.7.x} and
\texttt{source-map-support@0.5.y}. However, the latter depends on \texttt{source-map@0.6.z}, thus the build must include both versions of \texttt{source-map}. The ideal fix would update \texttt{source-map-support} to support \texttt{source-map@0.7.x}.

\subsection{\rqtwo}


\def\testRunningMatrix{{
{563,  7},  
{   38,127},  
}}

\def\testResultLabels{{"Pass","Fail"}} 
\def\numTestResultLabels{2} 
\def\testRunningMatrixScale{1.5} 

\begin{figure}
\begin{center}
\hspace{-40pt}
\begin{tikzpicture}[
    scale = \testRunningMatrixScale,
]

\tikzset{vertical label/.style={rotate=90,anchor=east}}   
\tikzset{diagonal label/.style={rotate=45,anchor=north east}}

\def\totSamples{0}
\foreach \y in {1,...,\numTestResultLabels}
{
    \foreach \x in {1,...,\numTestResultLabels}
    {
        \pgfmathparse{\testRunningMatrix[\y-1][\x-1]}
        \xdef\totSamples{\totSamples+\pgfmathresult}
    }
}
\pgfmathparse{\totSamples} \xdef\totSamples{\pgfmathresult}  

\foreach \y in {1,...,\numTestResultLabels} 
{
    \node [anchor=east] at (0.4,-\y) {\pgfmathparse{\testResultLabels[\y-1]}\pgfmathresult}; 
    
    \foreach \x in {1,...,\numTestResultLabels}  
    {
    \begin{scope}[shift={(\x,-\y)}]
        \def\mVal{\testRunningMatrix[\y-1][\x-1]} 
        \pgfmathtruncatemacro{\r}{\mVal}   %
        \pgfmathtruncatemacro{\p}{round(\r/\totSamples*100)}
        \coordinate (C) at (0,0);
        \ifthenelse{\p<50}{\def\txtcol{black}}{\def\txtcol{white}} 
        \node[
            draw,                 
            text=\txtcol,         
            align=center,         
            fill=black!\p,        
            minimum size=\testRunningMatrixScale*10mm,    
            inner sep=0,          
            ] (C) {\r\\\p\%};     
        \ifthenelse{\y=\numTestResultLabels}{
        \node [] at ($(C)-(0,0.75)$) 
        {\pgfmathparse{\testResultLabels[\x-1]}\pgfmathresult};}{}
    \end{scope}
    }
}
\coordinate (yaxis) at (-0.4,-0.75);  
\coordinate (xaxis) at (0.5+\numTestResultLabels/2, -\numTestResultLabels-1.1); 
\node [vertical label] at (yaxis) {\namenpm};
\node []               at (xaxis) {NPM};
\end{tikzpicture}
\end{center}
\caption{
    Results of running tests after solving dependencies with NPM and \namenpm{}. In total only 5\% of packages have a failing test with \namenpm{} but not with NPM. 
}
\label{figure:test_result_matrix}
\end{figure}

\begin{table}[ht]
    \centering
    \begin{tabular}{lrr}
        \hline
    Statistic   & NPM Pass \&       & NPM Pass \&       \\
                & \namenpm{} Pass   & \namenpm{} Fail   \\
        \hline
    Mean            &  489.36       &   58.90   \\ 
    STD             &  4699.01      &   266.87  \\ 
    Minimum         &  0.00         &   0.00    \\ 
    25th Perc.      &  0.00         &   1.50    \\ 
    50th Perc.      &  7.00         &   4.00    \\ 
    75th Perc.      &  39.00        &   10.50   \\ 
    Maximum         &  81582.00     &   1493.00 \\ 
        \hline
    \end{tabular}
\caption{
    Statistics of the number of executed tests per package in the top left and top right groups of \cref{figure:test_result_matrix}.
}
\label{table:group_stats}
\end{table}

Although a package should pass its test suite with any set of dependencies that satisfy all constraints, in practice 
tests may fail with alternate solutions due to under-constrained dependencies.
We identified test suites for 735 of the \datasetTopName{} packages. A test suite succeeds only if all tests pass.
All test suites succeed with both \namenpm{} and NPM on 77\% of packages, and fail for both on 17\%. There are 38 packages where \namenpm{} fails but NPM passes, and 7 packages where NPM fails and \namenpm{} succeeds (\Cref{figure:test_result_matrix}).

Test failures occurring slightly more often with \namenpm{}'s solutions are likely due to the fact that many of these
packages have already been solved and tested with NPM, so even if their dependencies are underconstrained,
at present NPM produces working solutions. Manual investigation suggests that the 7 packages that fail with NPM but succeed with \namenpm{} are likely due to flaky tests and missing development dependencies 
while the 38 packages that fail with \namenpm{} but succeed with NPM are due to those reasons in addition to under-constrained
dependencies.

Finally, to verify that packages which pass their tests with \namenpm{} are not doing so vacuously due to no or few tests, 
in \Cref{table:group_stats} we report statistics of the number of executed tests for 
packages in the group that pass with NPM and \namenpm{}, and in the group that pass with NPM and fail with \namenpm{}.
A two-sided Mann-Whitney U test indicates that there is no statistically significant difference between the two populations (p = 0.49).

\subsection{\rqthree}
The rightmost three columns of Table~\ref{table:failures} show the number of failures resolving dependencies on the \datasetTopName{} for NPM, along with each configuration of \namenpm that we evaluated.
On the \datasetTopName{} packages, NPM itself fails on \dataNumNPMFailures{} packages.
Many of these failures occur due to broken, optional peer-dependencies that \namenpm{} does not needlessly solve.\footnote{They are not necessary to build, but NPM attempts to solve for them even with the \texttt{-{}-omit-peer} flag.}
We run \namenpm{} in several configurations, and get \dataMinMinNPMFailures{}---\dataMaxMinNPMFailures{} failures when \emph{duplicate versions are permitted}.
Some failures occur across all configurations, e.g., one package requires macOS.
Most of our other failures are timeouts: we terminate Z3 after 10 minutes.
When duplicates are not permitted, we do get more failures due to unsatisfiable constraints, but these are expected (\cref{section:evaluation:tree_solving}).
Some users may prefer to have \namenpm{} fail when it cannot find a solution rather
than falling back to an unconstrained solution, as 
the latter may lead to subtle and hard-to-debug issues at runtime due to e.g. 
conflicting global variables in multiple versions of the same package.
When we permit duplicates like NPM, we find that \namenpm successfully builds \emph{more} packages than NPM itself, providing strong evidence that \namenpm can reliably be used as a drop-in replacement for NPM.

\subsection{\rqfour}

\begin{figure}
    \includegraphics[width=\columnwidth]{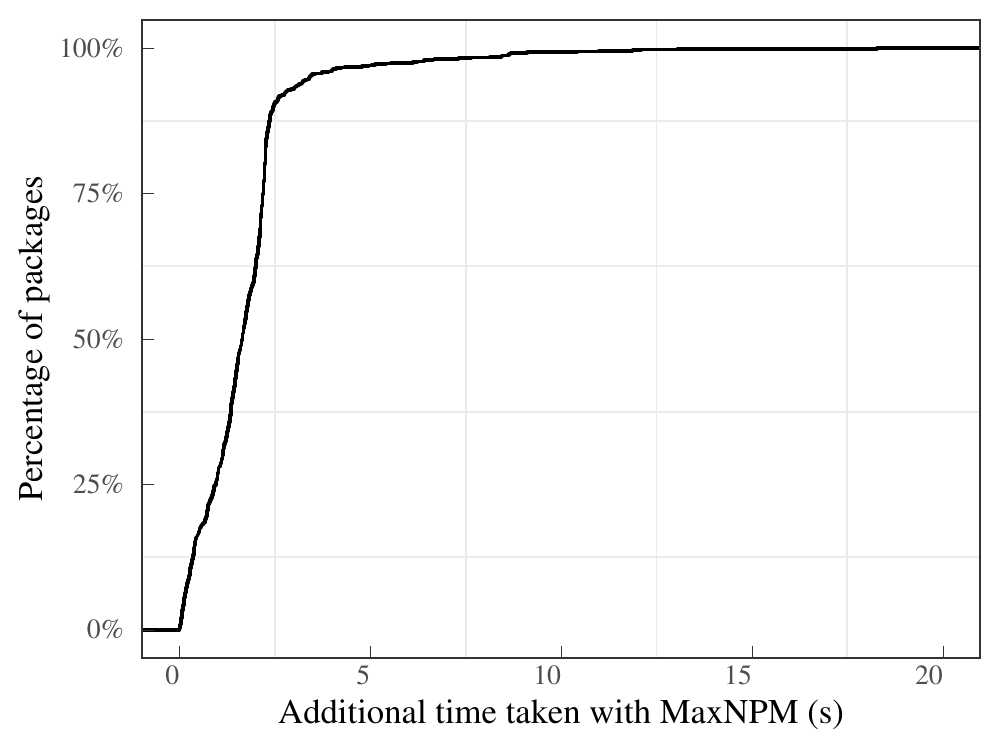}
    \caption{
        ECDF of the additional time taken by \namenpm{}
        to solve and install packages compared to NPM, ignoring timeouts and failures,
        with outliers ($> 20s$)  excluded.
        The outliers take up to \dataMaxSlowdown{} extra seconds, 
        but the mean and median slowdowns are only 
        \dataMeanSlowdown{} and \dataMedianSlowdown{}, respectively.
        In this experiment \namenpm{} was configured with
        NPM-style consistency, allowing cycles, and minimizing oldness first and
        then number of dependencies.
    }
    \label{figure:slowdown_no_outliers}
\end{figure}

On the \datasetTopName{} packages, we calculate how much \emph{additional time} \namenpm{} takes to solve dependencies over NPM.
We observe that the minimum slowdown is $\dataMinSlowdown{}$ (when \namenpm{} is faster than NPM), the 1st quartile is \dataFirstQuantileSlowdown, the median is \dataMedianSlowdown, the mean is \dataMeanSlowdown, the 3rd quartile is \dataThirdQuantileSlowdown{}, the max is \dataMaxSlowdown, and the standard deviation of the slowdown is \dataStdSlowdown{}. These absolute slowdowns are on top of the baseline of NPM, which takes 1.52s on average, and 1.34s at the median. 
We exclude timeouts from this analysis, we report those in \Cref{table:failures}.
As evidenced by the maximum and standard deviation, there are a few outliers where \namenpm{} takes substantially longer.
We also perform a paired Wilcoxon signed rank test and find that the slowdown is
statistically significant ($p<2.2 \times 10^{-16}$), with a moderate Cohen's d effect size of $d=0.27$.
\Cref{figure:slowdown_no_outliers} shows an ECDF of the absolute slowdown, but with outliers ($> 20$s) removed.
We conclude that while \namenpm{} does increase solving time, the increase is modest in the majority of cases, but there are a few outliers. This performance characteristic mirrors that of other SAT-solver based package managers, including production ones such as Conda~\cite{conda-slow}.

Excluding outliers, a significant portion of the overhead is serializing data between JavaScript (\namenpm) and Racket (\namebackend),
which could be improved by building the solver in JavaScript or using a more efficient serialization protocol. As for the outliers, one could implement a tool that first tries \namenpm{} but reverts to greedy solving after a timeout, at the expense of optimality.

\section{Related Work}
\label{related}

Van der Hoek et al. first discussed the idea of ``software release
management''~\cite{van1997software} for large numbers of independent packages in 1997,
and the first package managers for Linux distributions emerged at around the same
time~\cite{rpm,apt}. The version selection problem was first shown to be NP-complete and
encoded as a SAT and Constraint Programming (CP) problem by Di Cosmo et
al.~\cite{dicosmo:edos,mancinelli+:ase06-foss-distros} in 2005. This early work led to
the Mancoosi project, which developed the idea of a modular package manager with
customizable solvers~\cite{abate2012dependency,abate-2013-modular-package-manager}. This
work centers around the Common Upgradeability Description Format (CUDF), an input format
for front-end package managers to communicate with back-end solvers.

CUDF facilitated the development of solver {\it implementations} using Mixed-Integer
Linear Programming, Boolean Optimization, and Answer Set
Programming~\cite{michel+:lococo2010,argelich+:lococo2010,gebser+:2011-aspcud}, and many
modern Linux distributions have adopted CUDF-like approaches~\cite{abate2020dependency}.
OPIUM~\cite{tucker+:icse07-opium} examined the use of ILP with weights to minimize the
number of bytes downloaded or the total number of packages installed.

While package managers have their roots in Linux distributions, they have evolved
considerably since the early days. Modern language ecosystems have evolved their own
package managers~\cite{npm,pip,cargo,weizenbaum:pubgrub18}, with solver requirements
distinct from those of a traditional Linux distribution. Distribution package managers
typically manage only a single, global installation of each package, while language
package managers are geared more towards programmers and allow {\it multiple}
installations of the same software package.

For the most part, language ecosystems have avoided using complete solvers.
As we have found in our implementation, solvers are complex and interfacing with them 
effectively is more challenging compared to implementing a greedy algorithm.
Even on the Linux distribution side, so-called
\emph{functional} Linux distributions~\cite{dolstra+:lisa04,courtes-guix-2015}
eschew solving altogether, opting instead to focus on reproducible
configurations maintained by humans. Most programmers do not know how to use
solvers effectively, and fast, high-quality solver implementations do not exist
for new and especially interpreted languages. Moreover, package managers are now
fundamental to software ecosystems, and most language communities prefer to
write and maintain their core tooling in their own language.

Despite this, developers are starting to realize the need for completeness and
well defined dependency resolution semantics~\cite{abate2020dependency}. The
Python community, plagued by inconsistencies in resolutions done by PIP
has recently switched to a new resolver with a proper
solver~\cite{pip-new-resolver}. Dart now uses a custom CDCL SAT solver called
{\tt PubGrub}~\cite{weizenbaum:pubgrub18}, and Rust's Cargo~\cite{cargo} package
manager is moving towards this approach~\cite{pubgrub-rs}.
However, these solvers use ad hoc techniques baked into the implementations
to produce desirable solutions, such as exploring package versions sorted by version number.
These are not guaranteed to be optimal, and it is unclear how to add or modify objectives to these types of solvers.
In contrast, \namebackend{} makes two new innovations: \namebackend{} allows for a declarative specification
of multiple prioritized optimization objectives, and \namebackend{} changes the problem representation from
prior works' boolean-variable-per-dependency representation based on SAT solving to \namebackend{}'s symbolic
graph representation (\cref{sol-graph-synth}) based on SMT constraints.

Solvers themselves are becoming more accessible through tools like
Rosette~\cite{torlak2013growing}, which makes features of the
Z3~\cite{demoura+:tacas08} SMT solver accessible within regular
Racket~\cite{felleisen2015racket} code, and which we leverage to implement \namebackend{}. 
Spack~\cite{gamblin+:sc15} makes complex
constraints available in a Python DSL, and implements their semantics using
Answer Set Programming~\cite{gebser+:aicomm11,gamblin:fosdem20-concretizer}. APT
is moving towards using Z3 to implement more sophisticated dependency
semantics~\cite{klode:z3-apt}.

The goal of our work is to further separate concerns away from package manager
developers. \namebackend focuses on {\it consistency} criteria and
formalizes the {\it guarantees} that can be offered by package
solvers. NPM~\cite{npm}'s tree-based solver avoids the use of an
NP-complete solver by allowing multiple, potentially {\it inconsistent} versions
of the same package in a tree.
Tools like Yarn~\cite{yarn}, NPM's audit tool~\cite{npm-audit}, Dependabot~\cite{dependabot}, Snyk~\cite{snyk} and others~\cite{2021-maven-bloated,2021-the-used,2021-longi,2021-not-all} attempt to
answer various needs of developers by using ad hoc techniques separate from the solving phase, such
as deduplication via {\it hoisting}~\cite{maier:hoisting} (Yarn), or post hoc updating of dependencies (NPM's audit tool).
However, these tools run the risk of both correctness bugs and non-optimality in their custom algorithms. 
\namebackend{} provides the best of all these worlds. It combines the
flexibility of multi-version resolution algorithms with the guarantees of
complete package solvers and being able to reason about multiple optimization objectives that each speak
to a need of developers, while guaranteeing a {\it minimal} dependency graph.

\section{Discussion}
\label{discussion}


\subsection{NPM}
By modeling NPM in \namebackend{} and comparing their real-life behavior, we gained
valuable insight into NPM's behavior. As already explored in \cref{section:evaluation}, NPM is non-optimal,
and it is challenging to see how it could be optimal without implementing a full solver based approach such as
\namenpm{}. However, NPM does have some lower-hanging fruit that is easier to achieve and would benefit users.
First, NPM is not in fact \emph{complete}, in that there are situations where a satisfying solution exists, but NPM 
fails to find it. Most commonly, a version of a package depends on a dependency which does not exist, and NPM 
immediately bails out rather than backtracking. This could be implemented with simple backtracking without harming
the performance of solves which currently succeed.
In addition, \cref{npm-audit-could-be-better} identifies several shortcoming of the \texttt{npm audit fix} tool at the time
of our testing. We would suggest incorporating severity of vulnerabilities into the update logic, so that the tool
can decide trade-offs between different vulnerabilities. The tool is also unable to downgrade
dependencies to remove vulnerabilities, which would be a useful option to have, even if not enabled by default.




NPM also contains some subtle behavior regarding \emph{release tags}.
NPM version numbers may include release tags, such as \texttt{1.2.3-alpha1}, which \namenpm{} fully supports. 
In particular, a prerelease version can only satisfy a constraint
if a sub-term of the constraint with the same semver version also has a
prerelease.
For example, consider the following constraint (spaces indicate a conjunction):
\begin{verbatim}
>1.2.3-alpha.3 <1.5.2-alpha.8
\end{verbatim}
It may be obvious that versions
\texttt{1.2.3-alpha.7} and \texttt{1.5.2-alpha.6} do satisfy the constraint.
However, version \texttt{1.3.4} also satisfies the constraint, while
version \texttt{1.3.4-alpha.7} \emph{does not} satisfy the constraint.
A key feature of \namebackend{} which enabled us to easily implement this feature
is that the datatype of version numbers ($\domainVersions$) is not fixed to be e.g. a 3-tuple
of integers, but can be customized, which we leverage to correctly implement NPM's release tags.

\subsection{Future Work: Cargo, Maven and NuGet}
\label{section:future-work-other-languages}
Cargo has a notion of \emph{feature flags} which enable conditional compilation. To model this, versions may have the form
$(x, y, z, F)$, where $F$ is a set of selected features. 
To then build a Cargo-compatible dependency solver, we would need to 1)~define a constraint satisfaction predicate that ensures that the selected features are a superset of the requested feature, and 2)~an objective function that ensures that the minimal number of required features are enabled. In addition, Cargo has a more complex consistency function which allows
two versions to be co-installed if they are not semver-compatible. That is easily implemented in \namebackend{} as shown in
\cref{listing:consistency-comparison}.


\begin{figure}
  \centering
  \begin{minted}[
   linenos,
   numbersep=5pt,
   fontsize=\footnotesize,
   frame=lines,
   fontfamily=zi4,
   framesep=2mm]{racket}

(define (cargo-consistent v1 v2)
 (match `(,v1 ,v2)
   [`((0 0 ,z1) (0 0 ,z2))       #true]
   [`((0 ,y ,z1) (0 ,y ,z2))     (= z1 z2)]
   [`((0 ,y1 ,z1) (0 ,y2 ,z2))   #true]
   [`((x ,y1 ,z1) (x ,y2 ,z2))   (and (= y1 y2) (= z1 z2))]
   [_                            #true]))
\end{minted}
  \caption{A consistency function for Cargo.}
  \label{listing:consistency-comparison}
\end{figure}

Maven and NuGet resolve all dependencies to exactly one version for
each package. However, when conflicts arise, instead of backtracking to an older
version, they \emph{ignore all but the closest package constraint to the root}.
This is arguably unsound, but it is possible to model this behavior in \namebackend{} by treating constraints 
as soft constraints that are weighted by their distance from the root.

\subsection{Threats to Validity}
\label{threats}

\paragraph{External Validity}
\label{threats:external}
The projects that we used in our evaluation may not be representative of the entire ecosystem of NPM packages. 
We select the 1,000 most popular projects, and report performance as a distribution over this entire dataset, including a discussion of outliers. Given the number of projects that we used in our evaluation and their popularity, we believe that \namenpm{} is quite likely to be helpful for improving package management in real-world scenarios. We describe \namebackend{} as a unifying framework for implementing dependency solvers, however we only use \namebackend{} to implement \namenpm{}.
Future work should empirically validate \namebackend{}'s efficacy in other ecosystems.

\paragraph{Internal Validity}
\label{threats:internal}

\namenpm{}, \namebackend{}, and the tools that we build upon may have bugs that impact our results. To verify that differences
between NPM and \namenpm{} are not due to bugs in \namenpm{}, we carefully analyzed the cases where \namenpm{} and NPM diverged in their solution, and we walkthrough some example cases in \cref{section:evaluation}.
Additionally, we have carefully written a suite of unit tests for \namebackend{}.

\paragraph{Construct Validity}
We evaluate \namenpm{}'s relative performance to NPM when optimizing for several different objective criteria. 
However, it is possible that these criteria are not meaningful to developers.
For example, when comparing \namenpm{} and \texttt{npm audit fix} in reducing vulnerabilities, we use the aggregate vulnerability scores (CVSS) to rank the tools. However, in practice, these scores may not directly capture the true severity of a vulnerable dependency in the context of a particular application.
\namenpm{} does however allow for potential 
customization of constraints to fit the developer's needs. Future work should involve user studies, observing the direct impact of \namebackend{}-based solvers (including \namenpm{}, and implementations for other ecosystems) on developers.

\section{Conclusion}
\label{conclusion}

We present \namebackend, a semantics of dependency solving that we use to
highlight the essential features and variation within the package manager design space. We
use \namebackend to implement \namenpm, a drop-in replacement for NPM that
allows the user to customize dependency solving with a variety of global
objectives and consistency criteria.
Using \namenpm, developers can optimize dependency resolution to achieve goals that NPM is unable to, such as: reduce the presence of vulnerabilities, resolve newer packages and reduce bloat.
We evaluate \namenpm on the top 1,000 packages in the NPM ecosystem,
finding that our prototype introduces a median overhead of less than two seconds.
We find that \namenpm produces solutions with fewer dependencies and newer dependencies for many packages.
For future work, we hope to use
\namebackend to build new dependency solvers for other package managers as well.

\section{Data Availability}
Our artifact is available under a CC-BY-4.0
license~\cite{maxnpm-artifact} and consists of
\begin{inparaenum}
  \item the implementations of \namebackend{} and \namenpm{},
  \item the \datasetTopName{} and \datasetVulnName{} datasets, and
  \item scripts to reproduce our results.
\end{inparaenum}
All of our code is also available on GitHub~\cite{maxnpm-github}, and
\namenpm{} can be easily installed with \texttt{npm install -g maxnpm}.

\section*{Acknowledgments}

We thank Northeastern Research Computing, especially Greg Shomo,
for computing resources and technical support.

  {
    \footnotesize

    \ifthenelse{\equal{\formattype}{\formattypeIEEE}} {
      \bibliographystyle{IEEEtran}
    } {
      \bibliographystyle{ACM-Reference-Format}
    }

    \bibliography{bib/venues-short,bib/main}
  }

\end{document}